\documentclass[aps,prb,superscriptaddress,twocolumn,showpacs,a4paper,footinbib]{revtex4-1}
\usepackage[pdftex,breaklinks=true,bookmarks=false,colorlinks=true,citecolor=violet,urlcolor=blue!60!black,linkcolor=red]{hyperref}
\usepackage{amsfonts,amsmath}
\usepackage{graphicx}
\usepackage{color}
\usepackage{tikz}
\usetikzlibrary{shapes,decorations,calc}
\usepackage{braket}
\usepackage{verbatim}

\begin{document}
\title{Full counting statistics in the Haldane-Shastry chain}
\author{Jean-Marie St\'ephan}
\author{Frank Pollmann}
\affiliation{\mbox{Max Planck Institut f\"ur Physik komplexer Systeme, N\"othnitzer Str. 38, 01187, Dresden (Germany)}}
 \date{\today}
\begin{abstract}
We present analytical and numerical results regarding the magnetization full counting statistics (FCS) of a subsystem in the ground-state of the Haldane-Shastry chain.
 Exact Pfaffian expressions are derived for the cumulant generating function, as well as any observable diagonal in the spin basis.
  In the limit of large systems, the scaling of the FCS is found to be in agreement with the Luttinger liquid theory. The same techniques are also applied to inhomogeneous
  deformations of the chain. This introduces a certain amount of disorder in the system; however we show numerically that this is not sufficient to flow to the random singlet phase, that corresponds to $XXZ$ chains with uncorrelated bond disorder.
\end{abstract}

 \maketitle
 
 \section{Introduction}
 \subsection{Full counting statistics and the Haldane-Shastry chain}
 Fluctuations of macroscopic observables in quantum systems provide useful information about their underlying physical properties: 
 correlations, excitations, transport, to name a few. One example is that of charge fluctuations in mesoscopic systems such as (fractional) quantum Hall
 samples \cite{FQH_fluc1,FQH_fluc2}. From a theoretical perspective, the object of interest is the full distribution of transmitting charges, known as the full counting statistics (FCS).
 Such an object encodes the information about charge fluctuations -- that are sometimes experimentally accessible -- but also all higher order correlations. The FCS for a charge operator $Q$ is defined as
 \begin{equation}\label{eq:fcsdef}
  \chi(\lambda)=\Braket{e^{i\lambda Q}}=\sum_{m=0}^\infty \frac{(i\lambda)^m}{m!}\Braket{Q^m},
 \end{equation}
 where $\lambda$ is a counting parameter. 
It is nothing but the generating function for all the moments of the charge fluctuations. It is also extremely useful to consider the logarithm of $\chi$, which is a generating function for the cumulants\cite{}. The physically most important is the second cumulant, $\Braket{Q^2}_c=\Braket{Q^2}-\Braket{Q}^2$. All these concepts can also be applied to many body systems where a global conserved quantity (not necessarily charge) is measured in a subsystem. The study of such ``bipartite fluctuations'' has mainly been put forward in Refs.~\onlinecite{Yalefluc1,Yalefluc2}. 
 
 In the context of mesoscopic systems, Levitov and Lesovik found\cite{LevitovLesovik} a simple determinant formula for the FCS, that triggered an intense
  theoretical interest in the subject\cite{Meso}. The FCS turns out to also be an important tool in the study of the Fermi edge problem\cite{AbaninLevitov}, as well as cold atom systems\cite{Gritsev2006,Polkovnikov2006}. For physical
  setups that boil down to free fermionic problems, the FCS can always be expressed in determinant or Pfaffian form (see Ref.~\onlinecite{Klichfluc} for a general discussion). This result carries through
  for spin chains that can be mapped onto free fermions through a Jordan-Wigner transformation.
  
  Another motivation to study fluctuations lies in the relation with the entanglement entropy. For free fermions there is a precise correspondence between the two\cite{KlichLevitov,Yalefluc1,Yalefluc2}. This however does not survive when adding interactions (see e.g. Ref.~\onlinecite{Yalefluc3}). While both quantities diverge logarithmically at low energies for spin chains described by a Luttinger Liquid (LL) theory, the coefficient of the EE is controlled by the central charge in general, while the coefficient of the FCS is essentially the Luttinger parameter.
  
  Full counting statistics may also be used to track disorder \cite{Yalefluc1}. For example, it is well known that Heisenberg type chains with random bonds are effectively described by an infinite disorder critical point, the random singlet phase (RSP) \cite{FisherRandom}. In that case the second cumulant also diverges logarithmically (with a different coefficient), but so do all even order cumulants. Hence the FCS provides a simple way of distinguishing between the LL and RSP phases in spin chains. Fluctuations have also been used as a reliable way to locate many-body localization transitions\cite{LuitzLaflorencieAlet,Pollmannandothers}. 
  
  Despite all that,
  general analytical results are difficult to obtain in the presence of interactions. From a technical perspective, computing the FCS typically requires the knowledge of all correlation functions
  in a certain spatial region, which is a formidable task even for integrable systems. 
  We demonstrate here that these difficulties can be overcome in the Haldane-Shastry (HS) model, a chain with $SU(2)$ symmetry and long range interactions that still exhibits Luttinger liquid physics. 
  The HS Hamiltonian takes the peculiar form
  \begin{equation}\label{eq:hsham}
  H=\sum_{i \neq j=1}^L \frac{\mathbf{S}_i .\mathbf{S}_j}{\left(\frac{L}{\pi}\sin \frac{\pi (i-j)}{L}\right)^2},
  \end{equation}
for a chain of $L$ sites with periodic boundary conditions. 
From a low energy field theory perspective, this model lies in the same universality class
 as the Heisenberg chain, a Luttinger liquid at the self dual point, also known as the $SU(2)_1$ Wess-Zumino-Witten (WZW) conformal field theory (CFT). Such Hamiltonians and generalizations 
 have received renewed attention over the last few years, as they
 can be constructed in a rather systematic way by discretizing conformal field theories 
 \cite{CiracSierra,NielsenCiracSierra,Bondesansun1,TuNielsensun1,TuSierra}. Alternative derivations are also possible \cite{Greiter1,Greiter2}. Similar constructions may be applied to higher dimensions, and can be used to mimick chiral topological phases similar to fractional quantum Hall states in two dimension \cite{KalmeyerLaughlin,NielsenCiracSierra2}. In all cases, the corresponding states can be seen as matrix product states with an infinite bond dimension, hence they are often dubbed infinite matrix product states (IMPS)\cite{CiracSierra}.
 
 The purpose of this paper is to study the FCS in this system, as well as inhomogeneous generalizations of it. For the sake of simplicity, we will focus on ground state physics.
 A remarkable property of the HS chain is that its ground-state wave function
 takes a particularly simple Jastrow-type form, after mapping the spins onto hardcore particles (the down spins are then holes). The number of particles is fixed by the magnetization sector, for example at zero magnetization there are $N=L/2$ particles. In fractional quantum Hall language, this is a lattice discretization of the (bosonic) Laughlin $\nu=1/2$ state on a ring. For other values of $\nu$ these variational states are Luttinger liquids provided $\nu>1/4$ \cite{Shastrygas}.

 Alternatively, such states can be represented using a discrete lattice gas picture on a ring, with 2d Coulomb interactions between the particles.
  In the limit $L\to \infty$ with $N$ fixed, this gas coincides with the (continuous) Dyson gas\cite{Dysongas}, familiar in the study of random matrices \cite{Mehta,Forrester} and the Calogero-Sutherland model\cite{Calogero,Sutherland1,Sutherland2}. Correlations in this limit are exactly those of the circular ensemble for symplectic matrices; it has been long known that they are given by Pfaffians\cite{Mehta}. Similar Pfaffian formulas also hold in the discrete case\cite{MehtaMehta74}, and such formulas can be used to reconstruct the multispin correlation functions of the $S_j^z$. Here we will use such methods to compute exactly the FCS in Pfaffian form; this allows for large scale numerical computations as well as analytical asymptotic results. We will also provide a generalization to systems with open boundary conditions.

  As we shall see, these technical results can be generalized to a certain class of inhomogeneous states, where the particles need not be regularly spaced on the circle. Such states are still ``integrable'', in the sense that correlations and fluctuations can still be computed exactly. This makes it natural to consider the effect of disorder in such HS-type chains.
  While it is possible to disorder the $XX$ chain without breaking the free fermionic nature of the system, one conclusion of our study will be that it is not so easy for HS type chains. In particular all natural ``integrable'' inhomogeneous versions that we will study
   will still lie in the Luttinger liquid universality class, because of the heavily correlated nature of the disorder. 
\subsection{Organization of the manuscript}
The manuscript is organized as follows. We start by a detailed study of the clean case in section \ref{sec:clean}. We first recall the Luttinger liquid description
 of the HS chain and use it to compute the FCS in the continuum. We  then present exact determinant formulae for the fluctuations in a subsystem of arbitrary size, before analyzing their asymptotic behavior. The results are found to be in agreement
 with Luttinger liquid predictions. In particular, it is shown that only the second cumulant diverges logarithmically, a clear signature of a gaussian effective theory.
 
 In section \ref{sec:disorder} we focus on inhomogeneous versions of the the HS chain, Eq.~(\ref{eq:hsham}). It is shown that such inhomogeneous ``integrable'' modifications of the HS chain/ground-state still lie 
 in the LL universality class.
 We end up by a general discussion of all the results in section \ref{sec:conclusion}. Most technical details regarding the derivation of our formulas for correlations and fluctuations are gathered in the appendix~\ref{sec:app1}. 

 \section{Clean case}
 \label{sec:clean}
 \subsection{Relation with $SU(2)_1$ Wess-Zumino-Witten}
 Perhaps one of the most striking property of the HS chain is the exceedingly simple form of its energy spectrum. The $n$-th energy state is \cite{Haldane,Shastry}
 \begin{equation}
  E_n-E_0=\frac{2\pi v\, q_n}{L}
 \end{equation}
 where $v$ is a velocity, and the $q_n$ are rational numbers. Such a result is typical for spin chains described by a CFT \cite{Cardy_c,Affleck_c}, however there are usually finite-size effects, that produce higher order (typically in powers of $L$) corrections.
 Here the conformal spectrum is exact on the lattice. This property holds because the chain has an infinite dimensional Yangian symmetry\cite{Yangian}: in a sense the HS chain is as close as can be to a pure CFT on a lattice.
 
 The ground-state wave function also takes a simple form. Let us map each spin configuration onto hardcore particles positions $x_1,\ldots,x_N$ in the set $\{1,2,\ldots,L\}$. Each site is occupied by a particle if the spin is up, unoccupied if the spin is down. There are $N=L/2$ particles since the ground state lies in the sector with zero magnetization. 
  The ground state reads $\Ket{\Psi_{\rm gs}}=\sum_{\{x\}} \psi(\{x\})\ket{\{x\}}$, where the sum runs over all allowed positions of the particles, and the amplitude is, up to a sign
   \begin{equation}
  \label{eq:hsgroundstate}
  \psi(\{x\})=\psi(x_1,\ldots,x_N)=\frac{1}{Z_\beta}\prod_{i<j} \left(\sin \frac{\pi (x_i-x_j)}{L}\right)^{\beta/2},
 \end{equation}
 with $\beta=4$ (in FQH language $\beta=2/\nu$). In this paper we need not worry about various phase factors in the amplitudes, as we focus on the statistics of the magnetization, which is diagonal in the $S_j^z$ or particle basis. 
 It is also good to keep in mind that the case $\beta=2$ corresponds to the ground-state of another well known spin chain, the $XX$ spin chain.
 
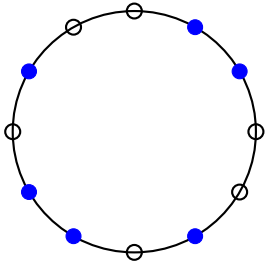
\begin{figure}[htbp]
   \begin{tikzpicture}[scale=0.8]
    \draw[thick] (0,0) circle (2cm);
    \foreach \i in {0,3,4,6,9,11}{
    \draw[thick] ({2*cos(30*\i)},{2*sin(30*\i)}) circle (3.5pt);
    }
     \foreach \i in {1,2,5,7,8,10}{
    \draw[color=blue,fill=blue] ({2*cos(30*\i)},{2*sin(30*\i)}) circle (3.5pt);
    }
   \end{tikzpicture}
   \caption{Representation of a particular amplitude $\psi(1,2,5,7,8,10)$ in the ground state wave function (\ref{eq:hsgroundstate}). Here the system is of size $L=12$, with $N=L/2=6$ particles (up spins). The particles at sites $1,2,5,7,8,10$ are shown as filled blue circles, and the holes at positions $0,3,4,6,9,11$ as empty circles.}
   \label{fig:particles_circle}
  \end{figure}

 Each amplitude in (\ref{eq:hsgroundstate}) may be represented as a gas of $N$ particles on a ring, as is shown in Fig.~\ref{fig:particles_circle}. Such a cartoon will be extremely useful in the remainder of the paper.
  
Here our aim is to focus on correlations, which \emph{do} have finite-size effects. For example the following  formula
 \begin{equation}
 \label{eq:correlator}
  \Braket{S_{x_1}^z\ldots S_{x_n}^z}=\underset{1\leq i,j\leq n}{\rm Pf} \left[a(x_i-x_j)\right]
  \underset{1\leq i,j\leq n}{\rm Pf} \left[d(x_i-x_j)\right]
 \end{equation}
 holds for the $n$-point function between spins in the $z$ basis. 
 Here the auxiliary functions $a$ and $d$ are given by 
\begin{eqnarray}\label{eq:a}
 a(x)&=&\sum_{k=1}^{L/2}\frac{\sin \frac{(2k-1)\pi x}{L}}{\pi(2k-1)},\\\label{eq:b}
 d(x)&=&\frac{(-1)^x}{\frac{2L}{\pi}\sin \frac{\pi x}{L}},
\end{eqnarray}
with the convention $d(0)=0$. Such a result is implicit in the random matrix literature (see in particular, Ref.~[\onlinecite{MehtaMehta74}]), however, we present a self-contained derivation  in appendix~\ref{sec:app1}. 
 
 Let us now discuss the form of this correlator. 
The second Pfaffian in (\ref{eq:correlator}) is exactly the $SU(2)_1$ CFT correlator for vertex operators, and the first encodes finite-size effects. For $n=2$ this reduces to the result of Refs.~[\onlinecite{BBG,NielsenCiracSierra}], $\braket{S_{x_1}^z S_{x_2}^z}=a(x_1-x_2)d(x_1-x_2)$, obtained with different methods.
 Note that such formulas can sometimes be useful to benchmark numerical techniques in long range systems \cite{LongRangeTimeMPS}, which are typically quite challenging.
\subsection{FCS for Luttinger liquids}
The generating function can be evaluated in the continuum limit using bosonization, as was done in Refs.~[\onlinecite{Yalefluc2,AbanovIvanovQian}]. For the sake of completeness, we recall here the result and the derivation. First, 
the HS chain is described in the continuum limit by the euclidean action
\begin{equation}\label{eq:llaction}
 S=\frac{1}{8\pi K}\int_0^L dx \int_{-\infty}^\infty d\tau\, (\nabla \varphi)^2+{\rm irr},
\end{equation}
where $\varphi=\varphi(x,\tau)$ lives on an infinite cylinder of circumference $L$. The field is compactified on a circle of radius one, $\varphi=\varphi+2\pi$. The Luttinger parameter is
$K=1/2$ for the HS chain, but we leave it unspecified for now. In the r.h.s of (\ref{eq:llaction}), irr denotes a set of irrelevant terms that do not affect the long-distance properties of the system. The magnetization operator in a subsystem of size $\ell$ is given by
\begin{equation}
 M_\ell=\sum_j S_j^z\sim\int_0^\ell \left(\rho(x)-\braket{\rho(x)}\right)dx,
\end{equation}
where $\rho(x)$ measures the particle density at position $x$. In bosonization language the density is the derivative of the field $\varphi$, $\rho(x)-\braket{\rho(x)}=\frac{1}{\pi}\partial_x \varphi$. Hence the generating function is given by 
\begin{eqnarray}
 \chi_\ell(\lambda)&=&\Braket{e^{i\lambda M_\ell}}\\\label{eq:beforewick}
 &=&\Braket{e^{i\frac{\lambda}{\pi} \left(\varphi(\ell)-\varphi(0)\right)}}\\\label{eq:afterwick}
 &=&e^{-\frac{\lambda^2}{\pi^2}\Braket{\left(\varphi(\ell)-\varphi(0)\right)^2}}.
\end{eqnarray}
(\ref{eq:afterwick}) follows from (\ref{eq:beforewick}) by applying Wick's theorem. Therefore, the calculation boils down to that of a two point function $\Braket{\varphi(\ell,0)\varphi(0,0)}$ on the cylinder,
 which is a standard CFT exercise \cite{BigYellowBook}.
We finally obtain\cite{Yalefluc2}
\begin{equation}\label{eq:fluc_ll}
 -\log \chi_\ell(\lambda)=K\frac{\lambda^2}{2\pi^2}\log\left(\frac{L}{\pi}\sin \frac{\pi \ell}{L}\right)+{\rm cst}(\lambda)+o(1).
\end{equation}
This result is valid in the limit $L \to \infty$, with fixed aspect ratio $\ell/L$, but also captures the $L\to \infty $, and only then $\ell \to \infty$ limit.
To the leading order, the cumulant FCS is proportional to $\lambda^2 \log \ell$.
Therefore only the second cumulant is diverging logarithmically, with a universal coefficient. This property reflects the fact that the most relevant part of the action (\ref{eq:llaction}) is purely gaussian.
The irrelevant operators in (\ref{eq:llaction}) contribute to $O(1)$ and lower order terms, so that all higher (even) order cumulants saturate
 to some finite non universal value as $\ell \to \infty$.
\subsection{Finite-size Pfaffian formula for the FCS}
\begin{figure}[htbp]
 \includegraphics[width=0.49\textwidth]{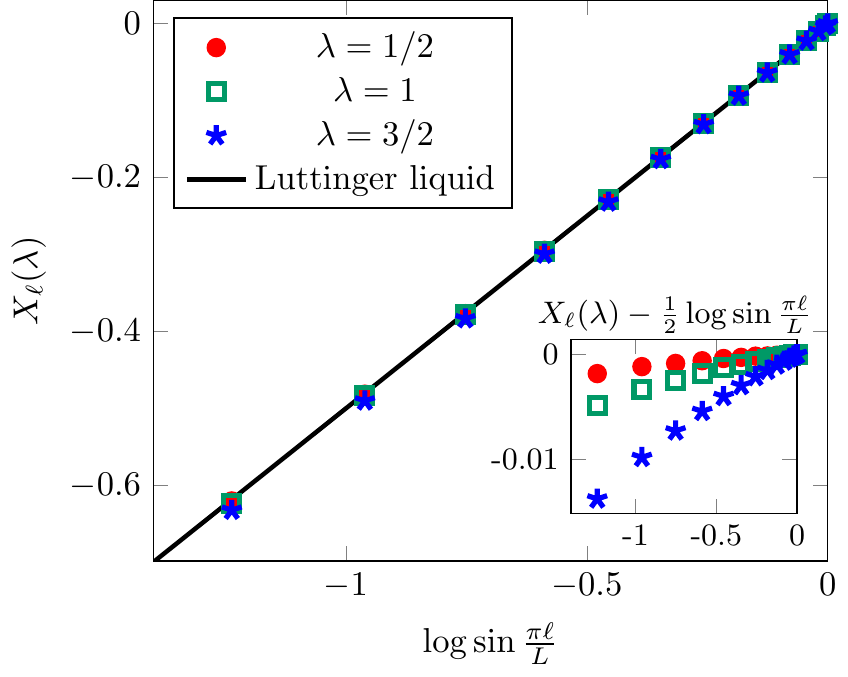}
 \caption{Convergence of $X_\ell(\lambda)$ to the universal scaling function of Eq.~(\ref{eq:scalingfunction}) with $K=1/2$ (black thick line). Numerical values for $L=4096$ and $\lambda=1/2,1,3/2$ are shown. The inset shows the difference with the Luttinger liquid prediction. As can be seen the finite size effects are small but become bigger as $\lambda$ is increased. This is due to the periodicity $\chi_\ell(\lambda+2\pi)=\chi(\lambda)$ of the FCS, which is not obeyed by the asymptotic expansion (\ref{eq:fluc_ll}). We refer to Ref.~\onlinecite{Abanovper} for discussion in the case of free fermions.}
 \label{fig:cfttests}
\end{figure}

Similarly to the case of free fermions, the FCS for the HS chain can also be obtained in a closed compact Pfaffian  form for all filling fractions. The technical details are gathered in appendix~\ref{sec:app1}. At half filling the result takes the elegant form
\begin{equation}\label{eq:latticefluc}
 \chi_\ell(\lambda)=\underset{1\leq i,j\leq \ell}{\rm Pf}
 \left(
 \begin{array}{ccc}
 2\sin \frac{\lambda}{2}\, a(i-j) && \cos \frac{\lambda}{2} \,\delta_{ij}\\\\
 -\cos \frac{\lambda}{2} \,\delta_{ij} && 2\sin \frac{\lambda}{2}\,d(i-j)
 \end{array}
 \right)
\end{equation}
for a subsystem of $\ell$ consecutive spins. Here $a$ and $d$ are given by Eqs.~(\ref{eq:a}) and (\ref{eq:b}) respectively. Recall that the Pfaffian of an antisymmetric matrix $K=(K_{ij})_{1\leq i,j \leq n}$ is defined as
\begin{equation}
 {\rm Pf}\, K=\frac{1}{2^n n!}\sum_{\sigma \in S_{2n}}(-1)^P K_{\sigma(1)\sigma(2)}\ldots
 K_{\sigma(2n-1)\sigma(2n)}
\end{equation}
where the sum runs over all permutations of $\{1,2,\ldots,2n\}$. The Pfaffian satisfies the important property $\left({\rm Pf} \,K\right)^2=\det K$. The expression (\ref{eq:latticefluc}) is extremely useful, both analytically and numerically. Let us first discuss the second cumulant $C_2$. We have
 \begin{eqnarray}
 C_2&=&-\left.\frac{d^2\log \chi_\ell}{d\lambda^2}\right|_{\lambda=0}\\
  &=&\frac{\ell}{4}+2\sum_{j=1}^\ell (\ell-j) a(j)d(j)
 \end{eqnarray}
The asymptotics ($L\to \infty$, $\ell/L$ fixed) converges to 
\begin{equation}
 C_2=\frac{1}{2\pi^2}\log \left(\frac{L}{\pi}\sin \frac{\pi \ell}{L}\right),
\end{equation}
compatible with the cumulant obtained from the bozonization result (\ref{eq:fluc_ll}).
 While we were not able to derive the full generating function (\ref{eq:fluc_ll}) from the lattice result (\ref{eq:latticefluc}), it is straightforward to evaluate the FCS numerically for various values of $\lambda$
 and very large system sizes. Assuming (\ref{eq:fluc_ll}), the normalized FCS $X_\ell(\lambda)=-(\pi^2/\lambda^2)\log [\chi_\ell(\lambda)/\chi_{L/2}(\lambda)]$ converges to a universal scaling function
 \begin{equation}\label{eq:scalingfunction}
  X_\ell(\lambda)= K \log \left(\sin \frac{\pi \ell}{L}\right)+o(1)
 \end{equation}
 that becomes independent on $\lambda$.
Some numerical results are presented in Fig.~\ref{fig:cfttests} for a chain of length $L=4096$, and confirm the validity of (\ref{eq:fluc_ll}) in the HS chain.

We finally comment on the case of open chains\cite{SimonsAltshuler,BernardPasquierSerban}, which have similar behavior. We consider an open HS chain of length $L$, and measure the fluctuations in a subsystem that consists of the first $\ell$ spins on the left. The predictions of the Luttinger liquid theory are reduced by a factor one-half in this case, so we can make the substitution $K\to K/2$ in (\ref{eq:fluc_ll}). Correlations may also be computed on the lattice in this case. For example the two point function has been recently derived\cite{TuSierra} using the Null vector construction mentioned in the introduction. Our method can also be applied to derive all (diagonal) correlations, as well as the FCS. Once again these are given by Pfaffian, we refer to the appendix \ref{sec:open} for the expressions and the derivation. With these formulae at hand we also checked that the predictions of Luttinger liquid also hold in the open case (not shown).

\subsection{Infinite system and block Toeplitz Pfaffian}
We now specialize to the limit $L\to \infty$, but at any filling fraction $\rho=N/L$. We get the block Pfaffian $\chi_\ell(\lambda)={\rm Pf}([g]_{i-j})$, 
 where 
 \begin{equation}
  [g]_k=\left(
 \begin{array}{ccc}
 \sin \left(\frac{\lambda}{2}\right)\frac{\cos 2\pi \rho k}{\pi k}&&e^{-i\frac{\lambda}{2}}\delta_{k0}+i\sin \left(\frac{\lambda}{2}\right)\frac{\sin 2\pi \rho k}{\pi k}\\\\
 -(\ldots)&&\sin \left(\frac{\lambda}{2}\right){\rm Si}(2\pi \rho k)
 \end{array}
  \right)
 \end{equation}
 $-(\ldots)$ means that the two by two matrix is antisymmetric. 
 ${\rm Si}(x)=\int_0^x dt\; t^{-1}\sin t$ denotes the Sine Integral. 
 Such determinants (Pfaffians) whose matrix elements only depend on $i-j$ are called Toeplitz determinants (Pfaffians). 
 Here the $[g]_k$ can be interpreted as the $k$-th Fourier coefficient\footnote{We choose the convention $[g]_k=\frac{1}{2\pi}\int_{-\pi}^{\pi}e^{-ik\theta} g(\theta)$ and $g(\theta)=\sum_{k\in \mathbb{Z}} [g]_k e^{ik\theta}$} of the $2\times 2$ matrix function
  \begin{equation}
  g(\theta)=\left(
 \begin{array}{ccc}
  \sin \left(\frac{\lambda}{2}\right)\theta&&\cos \frac{\lambda}{2}\\\\
  -\cos \frac{\lambda}{2}&& \sin \frac{\lambda}{2} \,{\rm pv}\left(\frac{1}{\theta}\right)
 \end{array}
 \right)
 \end{equation}
 for $|\theta|<2\pi \rho$, and 
 \begin{equation}
  g(\theta)=\left(
 \begin{array}{ccc}
  \sin \left(\frac{\lambda}{2}\right)(\theta-\pi \,{\rm sgn}\, \theta)&&e^{-i\frac{\lambda}{2}}\\\\
  -e^{-i\frac{\lambda}{2}}&& 0
 \end{array}
 \right)
 \end{equation}
 otherwise. ${\rm pv}(\frac{1}{\theta})=\frac{1}{2}\left(\frac{1}{\theta+i 0^+}+\frac{1}{\theta+i 0^{-}}\right)$ denotes the Cauchy principal value. 

We are interested in the asymptotics of $\log \chi_\ell(\lambda)$, when $\ell \to \infty$. Such questions have been widely studied in the mathematical literature\cite{Szego}, in part motivated by the celebrated Ising spontaneous magnetization problem (see Ref.~\onlinecite{IsingToeplitz} and references therein).
In the block matrix case fewer general theorems are available, and these typically require regularity assumptions that are not satisfied here. Ignoring this problem and applying nevertheless a formula of Widom's \cite{Widom1,Widom2}, the leading $\ell$ term is given by
\begin{eqnarray}\nonumber
\log \chi_\ell(\lambda)&\;\sim\;&
\frac{\ell}{2\pi} \int_{-\pi}^\pi \,d\theta \log \det g(\theta)\\
&\;\sim\;&i\lambda\ell \left(\rho/2-1/4\right)
\end{eqnarray}
This result is compatible with the naive argument 
\begin{equation}
 \braket{e^{i\lambda\sum_j S_j^z}}\approx e^{i\lambda \braket{\sum_j S_j^z}}=e^{i\lambda (\rho-1/2)\ell/2}
\end{equation}
for large $\ell$.
We also observe numerically
\begin{equation}
 \log \chi_\ell(\lambda)=i\lambda \left(\rho-1/2\right)\frac{\ell}{2}-\frac{\lambda^2}{4\pi^2}\log \ell+O(\ell^0),
\end{equation}
which is the expected result from bosonization. As emphasized before the coefficient of the logarithmic divergence is universal.

 \section{Inhomogeneous chains of the Haldane-Shastry type}
 \label{sec:disorder}
 \subsection{Moving the particles on the circle}
  \begin{figure}[htbp]
   \begin{tikzpicture}[scale=0.8]
    \draw[very thick] (0,0) circle (2cm);
    \foreach \i in {-0.13,3.18,3.81,5.77,9.12,10.97}{
    \draw[very thick] ({2*cos(30*\i)},{2*sin(30*\i)}) circle (3.5pt);
    }
     \foreach \i in {0.75,1.9,5.17,6.97,8.03,10.01}{
    \draw[color=blue,fill=blue] ({2*cos(30*\i)},{2*sin(30*\i)}) circle (3.5pt);
    }
    \draw (0,-3) node {(a) Particles};
    \begin{scope}[xshift=6cm]
      \draw[very thick] (0,0) circle (2cm);
    \foreach \i in {-0.13,3.18,3.81,5.77,9.12,10.97}{
    \draw[very thick,black] ({2*cos(30*\i)},{2*sin(30*\i)}) circle (3.5pt);
    \draw[<-,black,line width=2pt]  ({2*cos(30*\i)},{2*sin(30*\i)-0.4}) -- ({2*cos(30*\i)},{2*sin(30*\i)+0.4});
    }
     \foreach \i in {0.75,1.9,5.17,6.97,8.03,10.01}{
   \draw[very thick,blue] ({2*cos(30*\i)},{2*sin(30*\i)}) circle (3.5pt);
    \draw[->,blue,line width=2pt]  ({2*cos(30*\i)},{2*sin(30*\i)-0.4}) -- ({2*cos(30*\i)},{2*sin(30*\i)+0.4});
    }
    \draw (0,-3) node {(b) Spins};
    \end{scope}
   \end{tikzpicture}
   \caption{(a) Representation of a particular amplitude $\psi(1,2,5,7,8,10)$ in (\ref{eq:partdis}), in a inhomogeneous system of size $L=12$, $N=L/2=6$. (b) The analogous configuration in (\ref{eq:spinsdis}) may be obtained by assigning a $+$ spin to the particles (filled in blue) and a $-$ spin to the holes.}
   \label{fig:particles_circledis}
  \end{figure}
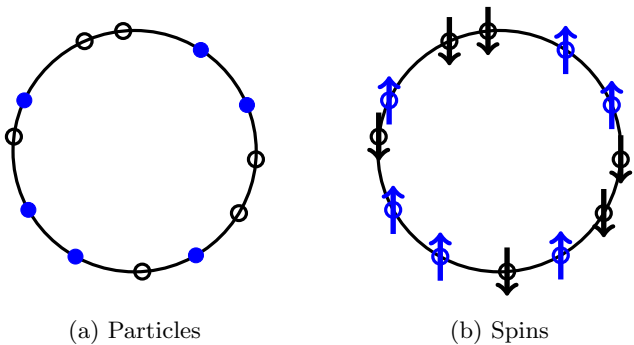
 Our exact formulae for the FCS and fluctuations can be generalized to any discrete set of allowed positions on the circle (see Eqs.~(\ref{eq:R},\ref{eq:f},\ref{eq:apq}) in the appendix). This makes it natural to consider the analogous wave function where the allowed positions on the circle are not regularly spaced anymore. Such inhomogeneities are a simple way of introducing some disorder into the system, as was pointed out in Ref.~\onlinecite{CiracSierra}. To be more precise, we consider the two classes of variational states
 \begin{equation}\label{eq:partdis}
  \psi(x_1,\ldots,x_N)=\frac{1}{Z_\beta}\prod_{1\leq j<k\leq N}\left|e^{i\theta_{x_j}}-e^{i\theta_{x_k}}\right|^{\frac{\beta}{2}},
 \end{equation}
and 
\begin{equation}\label{eq:spinsdis}
 \tilde{\psi}(\sigma_1,\ldots,\sigma_L)=\frac{1}{\tilde{Z}_{\beta}} \prod_{1\leq j<k\leq L}\left|e^{i\theta_j}-e^{i\theta_k}\right|^{\frac{\beta}{8}\sigma_j \sigma_k},
\end{equation}
but where the angles $\theta_j$ are no longer regularly spaced as ${\rm integer}\times \frac{2\pi}{L}$. As before we mainly focus on $\beta=4$.
The state (\ref{eq:partdis}) is pictured in Fig.~\ref{fig:particles_circledis}(a); the physical degrees of freedom are the particle positions. The other, (\ref{eq:spinsdis}), is slightly different, as all sites are now occupied by spins that can be either up ($\sigma_j=+1$) or down ($\sigma_j=-1$). Of course, we are still in the zero-magnetization sector, so that there are as many ($L/2$) spins up as spins down. This is represented in Fig.~\ref{fig:particles_circledis}(b).
Once again we only study diagonal observables in the $z$ (or particle) basis, so we need not be too careful about 
the respective phases of the amplitudes; they will be canceled when evaluating expectation values.
 
 In a clean system both states (\ref{eq:partdis}) and (\ref{eq:spinsdis}) turn out to be identical\cite{CiracSierra}, as can be seen by mapping the up spins in (\ref{eq:spinsdis}) to particles, the down spins to holes, and performing some algebra. This equality holds only when the $e^{i\theta_j}$ are $L-$th roots of unity, and so breaks down in the inhomogeneous case. Therefore we study them separately, even though they are quite similar. To avoid any confusion we dub (\ref{eq:partdis}) the inhomogeneous (bosonic) Laughlin state. The state (\ref{eq:spinsdis}) can be shown to be the exact ground-state of the following generalization of the HS Hamiltonian:
 
\begin{equation}\label{eq:hsgen}
 H=-\sum_{j\neq k} \,t_{jk}\, \mathbf{S}_j.\mathbf{S}_k,
\end{equation}
where
\begin{equation}\label{eq:tjk}
 t_{jk}= \frac{z_j z_k}{(z_j-z_k)^2}+\frac{w_{jk}(\alpha_j-\alpha_k)}{12},
\end{equation}
and $z_j=e^{i\theta_j}$, $w_{jk}=\frac{z_j+z_k}{z_j-z_k}=-i\cot\left(\frac{\theta_j-\theta_k}{2}\right)$ and $\alpha_j=\sum_{j\neq k}w_{jk}$. Despite the use of complex numbers, one can check that the couplings $t_{ij}$ in the Hamiltonian are all real.

\begin{figure*}[htbp]
 \includegraphics[width=0.6\textwidth]{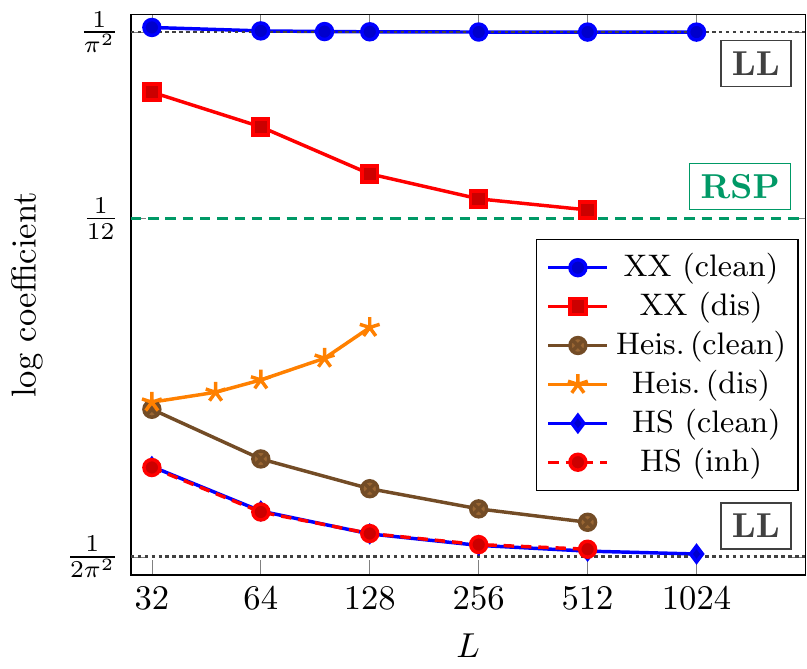}
  \caption{Coefficient of the logarithm in the bipartite fluctuations $C_2(L)$ for several chains cut into two halves. Each data point is obtained by computing $[C_2(L)-C_2(L/2)]/\log 2$ for a given $L$. The horizontal axis is shown in logarithmic scale. Blue circles represent the clean $XX$ chain, red squares the disordered $XX$ chain ($10^5$ disorder realizations), braun cross-circles the clean Heisenberg chain, orange stars the disordered Heisenberg chain ($10^3$ realizations), blue lozenges the clean HS chain, and dashed red circles the inhomogeneous HS chain ($10^5$ realizations). As can be seen the inhomogeneous HS chain still follows the Luttinger liquid prediction, contrary to the disordered XX and Heisenberg chains, that flow to the RSP phase (shown in dashed green). For the non clean points we choose $\delta=0.5$.}
 \label{fig:bfdis}
\end{figure*}
From now on we revisit the following problem \cite{CiracSierra}. We treat the angles $\theta_j$ as random variables
\begin{equation}
 \theta_j=\frac{2\pi}{L}\left(j+\delta_j\right),
\end{equation}
where the $\delta_j$ are uniformly distributed in $[-\delta;\delta]$. Since the set of couplings in the Hamiltonian (\ref{eq:hsgen}) are uniquely determined by the set of angles $\{\theta_j\}$, we are now effectively studying a Haldane-Shastry-type model with random bonds. Such random bonds spin chains have been widely studied in the case of nearest neighbor hoppings. For example, the $XXZ$ spin chain with random bonds and $-1/2<\Delta\leq 1$
\begin{equation}
 H=\sum_{j} t_j\left(S_j^x S_{j+1}^x+S_j^y S_{j+1}^y+\Delta S_j^z S_{j+1}^z\right)
\end{equation}
is known \cite{FisherRandom} to flow to a random singlet phase (RSP) if the $t_j$ are independent, and their distribution sufficiently regular. In the following we will only consider box distributions in the range $[1-\delta;1+\delta]$. The important point is that provided these two conditions are fulfilled, any amount of disorder will destabilize the Luttinger liquid. Note that despite being essentially localized, the random singlet phase still has critical correlations  when averaged over the disorder (we refer to Ref.~[\onlinecite{RefaelMoore2}] for a review). 

The question we wish to address here is the following: are the inhomogeneous states (\ref{eq:partdis},\ref{eq:spinsdis}) in the random singlet phase or still in the Luttinger liquid universality class?
The FCS offers a simple way to distinguish between the two phases. Indeed it also diverges logarithmically in the random singlet phase, but with a different prefactor \cite{Klichfluc}:
\begin{equation}\label{eq:fcsdis}
 \overline{\log \chi_\ell(\lambda)}=\frac{1}{3}\log \left(\cos \frac{\lambda}{2}\right)\log \ell+O(1),
\end{equation}
where $\bar{A}$ denotes the disorder averaged observable in the ground-state. 
Since the full finite size dependence is known only approximately\cite{Fagotti} for the RSP, we will consider only the simple situation of a periodic chain cut into two equal halves, which has the scaling shown in Eq.~(\ref{eq:fcsdis}), with $\ell=L/2$. Typically, checking formula (\ref{eq:fcsdis}) requires to compute the FCS of a large system for a particular disorder configuration, and repeating this procedure many times to access a truly disorder-averaged quantity. For most systems, especially with long range interaction, this is computationally extremely costly. The states (\ref{eq:partdis},\ref{eq:spinsdis}) are a notable exception, the other being of course the case of free fermions. We use it to answer our question numerically below, starting with the second cumulant, which is simplest.
\subsection{Second cumulant}\label{sec:secondcum}
The expected scaling of the second cumulant in our setup can be summarized by the two formulas
\begin{eqnarray}
 \label{eq:llc2}
 C_2^{\rm (LL)}&=&\frac{K}{\pi^2}\log L+O(1),\\
 \overline{C_2}^{\rm (RSP)}&=&\frac{1}{12}\log L +O(1),
 \label{eq:rspc2}
\end{eqnarray}
where LL stands for Luttinger liquid, and RSP stands for random singlet phase. Here $K$ is the Luttinger parameter. $K=1$ for free fermions, and $K=1/2$ for the HS chain. In the RSP phase there is no free parameter. We mainly consider the two states (\ref{eq:partdis},\ref{eq:spinsdis}). The second cumulant is evaluated using the results in the appendix \ref{app:correlations} for the state (\ref{eq:partdis}), and using the method of Ref.~[\onlinecite{NielsenCiracSierra}] for the state (\ref{eq:spinsdis}). 
For comparison, we also show data for the XX chain with and without disorder, as well as the Heisenberg chain, with and without disorder. For the former we use the free fermions structure to access very large system sizes with a large number of disorder ralizations; the computations for the latter are performed using the density matrix renormalization group (DMRG) \cite{White} method. 
 
An extraction of the coefficient of the logarithm in Eqs.~(\ref{eq:llc2},\ref{eq:rspc2}) is shown in Fig.~\ref{fig:bfdis} for all these models, and $\delta=0.5$.  
In the free fermions case, the slope is clearly different with and without disorder: it goes from $K/\pi^2=1/\pi^2\simeq 0.1013$ to the RSP value $1/12\simeq0.0833$. No such thing happens in the inhomogeneous HS states: the slope remains the same within our numerical accuracy, and the disorder only seems to affect the subleading term of order one. Given the very large system sizes considered, and the large number of disorder realizations, this data is strong evidence that the two inhomogeneous states are still in the Luttinger liquid universality class. For comparison, some data for the random Heisenberg chain is shown. The smaller accessible system sizes and statistical resolution makes it difficult to show convergence to the RSP value. The behavior we find is however consistent with a convergence to the RSP prediction. 
We also checked that all these behaviors remains true for any value $\delta<1$ of the disorder strength. 
\subsection{Full counting statistics}\label{sec:fullfcs}
\begin{figure}[htbp]
\includegraphics[width=0.48\textwidth]{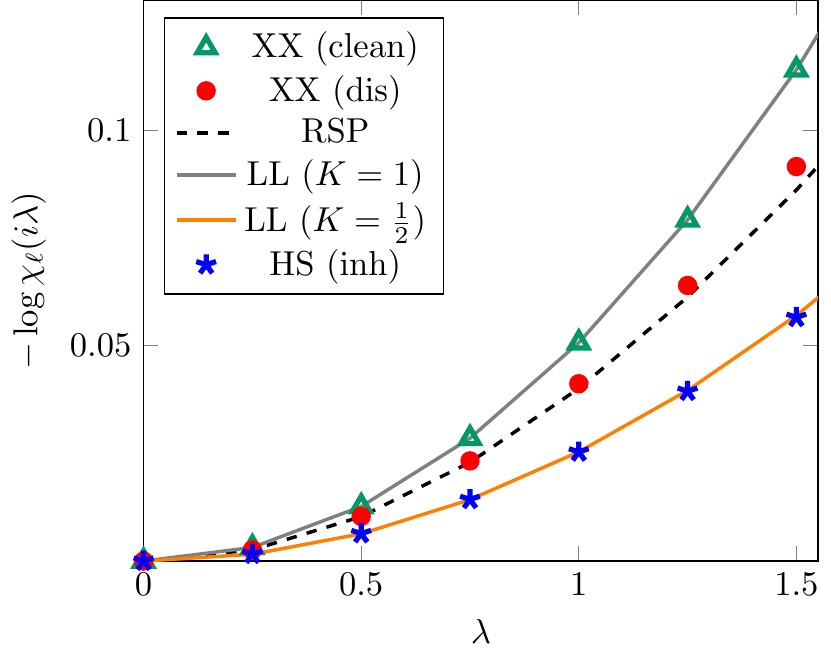}
\caption{Coefficient of the logarithm for the FCS generating function $\log \chi(i\lambda)$. Data for the clean $XX$ chain (green triangles), the disordered $XX$ chain (red circles), the inhomogeneous Laughlin state (blue stars). Predictions for the Luttinger liquid and RSP universality class are shown for comparison. As can be seen the bond disordered $XX$ curve flows to the RSP prediction, while the inhomogeneous Laughlin does not.}
 \label{fig:fcsdis}
\end{figure}
To confirm the results of the previous subsection, we turn our attention to the full generating function (\ref{eq:fcsdef}), still for a subsystem of $L/2$ consecutive spins in a periodic chain of length $L$. We use the same procedure as before, but perform fits to extract the coefficient of the logarithm. For each value of $\lambda$ we fit $\log \chi(i\lambda)$ or $\overline{\log \chi(i\lambda)}$ to $a\log L+b$ for $L=256,384,512$, and extract the resulting coefficient $a$. The reason we choose $i\lambda$ instead of $\lambda$ is mainly convenience, as it enhances the difference between clean an disordered system. The predictions can be deduced by just plugging $i\lambda$ instead of $\lambda$ in formula (\ref{eq:fluc_ll}) and (\ref{eq:fcsdis}). The expected behavior in the two phases is given by
\begin{eqnarray}\label{eq:fcsllbis}
  \log \chi(i\lambda)^{\rm (LL)}&=&\frac{K\lambda^2}{2\pi^2}\log L+O(1),\\\label{eq:fcsrspbis}
 \overline{\log \chi(i\lambda)}^{\rm (RSP)}&=&\frac{1}{3}\log \cosh \left(\frac{\lambda}{2}\right)\log L +O(1).
\end{eqnarray}
Our numerical results are shown in Fig.~\ref{fig:fcsdis}, and show essentially the same behavior as in the previous subsection. The clean XX chain follows the LL prediction, whereas the XX chain with random bonds behaves as expected in the RSP phase. Once again, the inhomogeneous Laughlin state behaves as expected for a Luttinger liquid. Note that we did not consider the inhomogeneous spin systems here, as the calculation of the full generating function is more complicated. However, we expect it to behave in the same way as its particle counterpart. 
\subsection{Interpretation}
\begin{figure}[htbp]
 \includegraphics[width=0.48\textwidth]{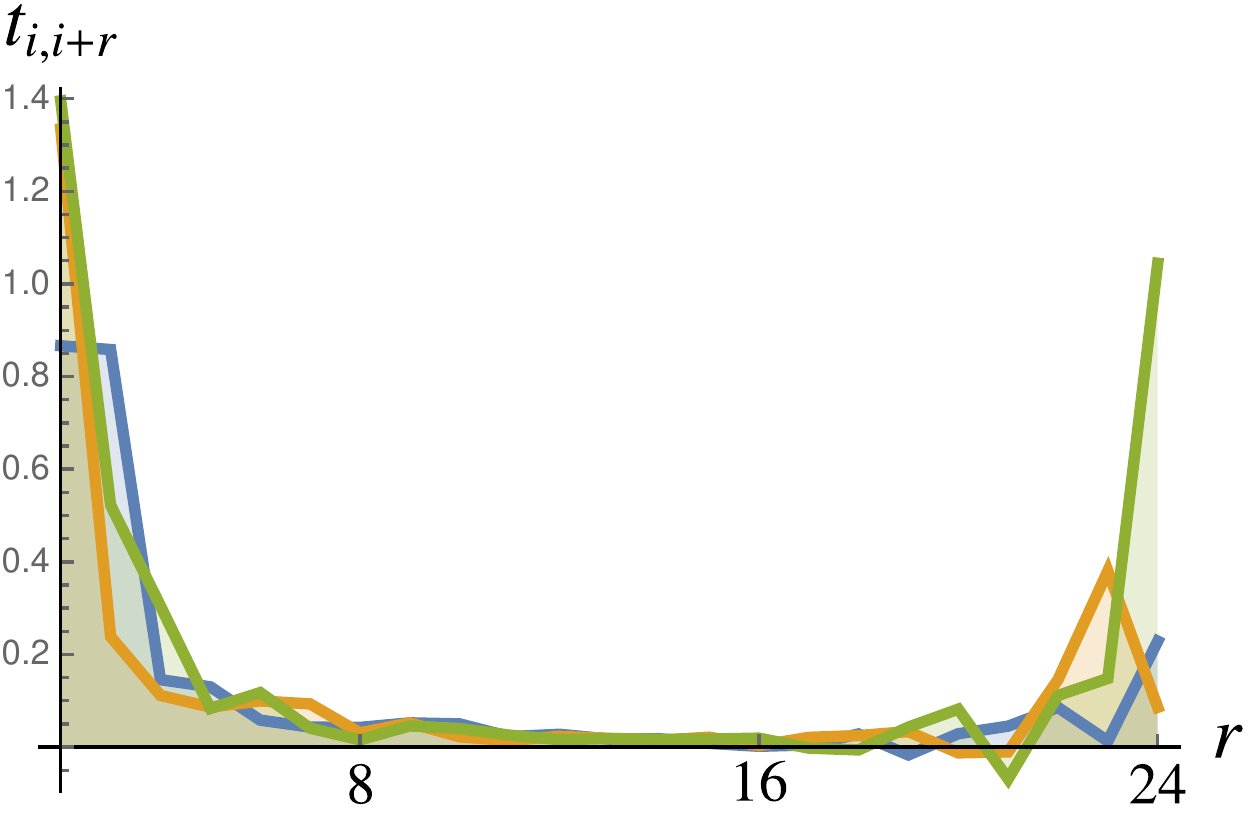}
 \caption{Bond amplitudes $t_{i,i+r}$ in (\ref{eq:hsgen}) as a function of $r$, for a chain of size $L=24$ and for three values of $i$. $i=1$ is shown in blue, $i=4$ in green, and $i=7$ in orange. The disorder strength is $\delta=0.5$.}
 \label{fig:correlated_disorder}
\end{figure}
Let us now try to interpret the numerical results of Secs.~\ref{sec:secondcum}, \ref{sec:fullfcs}.
As we have already mentioned, for spin chains with finite-range interactions (e.g. XXZ, J1-J2, etc.), any finite amount of (bond) disorder is sufficient to destabilize the Luttinger liquid, and produce a flow to the RSP. Such a phenomenon may be understood with a simple real space renormalization group picture (see Ref.~[\onlinecite{FisherRandom}]): what one does is identify the strongest bond in the chain, and form a spin singlet between the two lattice sites involved. These two are then effectively removed from the chain, and one computes the new coupling between the two sites that become nearest neighbors in the process. This is done using (second order) perturbation theory. The process is then repeated several times; after many RG steps what one finds is a collection of spin singlets, some of which are long range. This explains the critical correlations, when averaged over disorder.

One obvious issue with our class of variational states is that they correspond to Hamiltonians with long range interactions, and tracking this analytically using the RG method mentioned above is not straightforward. It is then a priori not obvious whether or not such systems can be driven to the RSP phase. Presumably the physics might also depend on the exponent governing the decay of the couplings in the Hamiltonian. We note that there are only a few studies of disorder in systems with long range interaction in the literature (see however, Ref.~\onlinecite{Boulder}). A systematic study of the interplay between disorder and the range of interactions goes beyond the scope of the present work, and is left as an important open problem.

There are however cases, even with finite range interactions, where the Luttinger liquid does not flow to the RSP. This can happen when the disorder is correlated, as was pointed out in Ref.~[\onlinecite{Laflorencie2}] for locally correlated disorder. This observation has been later confirmed in other systems (see e.g. Ref.~[\onlinecite{GetelinaAlcarazHoyos}]). Here we argue that the bond disorder in the Hamiltonian (\ref{eq:hsgen}) is, in fact, \emph{highly} correlated, even though the angles $\theta_j$ in the variational states (\ref{eq:partdis},\ref{eq:spinsdis}) are by definition independent and identically distributed. The reason is the complicated non linear mapping between the angles $\theta_j$, and the couplings $t_{jk}$ in Eq.~(\ref{eq:tjk}) which introduces correlations. To illustrate this we show in Fig.~\ref{fig:correlated_disorder} a few examples of the couplings $t_{jk}$ for one particular realization of the disorder. Another observation is that the couplings are not even positive anymore. 

\begin{figure}[htbp]
 \includegraphics[width=0.48\textwidth]{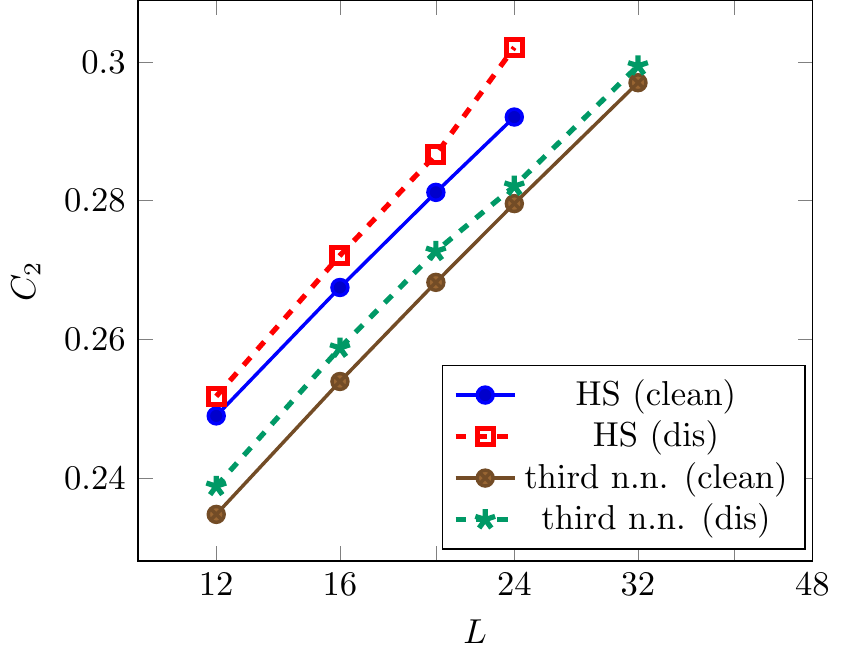}
 \caption{Second cumulant for the fluctuations in the ground state of the Hamiltonian (\ref{eq:hlongrange}) with disorder strength $\delta=0.5$ (red squares). The clean case is also shown for comparison (blue circles). We also present data for a truncated version where all couplings further than next-next nearest neighbors are set to zero for clean (brown circles) and disordered (green stars). The last two data sets are shifted by a constant offset $-0.015$ to improve readability. Each data point is averaged over at least $10^3$ realizations of the disorder, and the ground state found with DMRG for each realization.}
 \label{fig:dis_LR}
\end{figure}

In case of uncorrelated hoppings it becomes very difficult to perform numerical computations for large systems using DMRG. To illustrate this we considered the Hamiltonian
\begin{equation}\label{eq:hlongrange}
 H=\sum_{i\neq j} \frac{t_{ij}\,{\bf S}_i.{\bf S}_j}{\sin^2 \frac{\pi (i-j)}{L}} 
\end{equation}
where the $t_{ij}$ are again independent random numbers uniformly distributed in $[1-\delta,1+\delta]$. Some numerical results are shown in Fig.~\ref{fig:dis_LR} for system sizes up to $L=24$, and show a slight increase in the fluctuations compared to the clean HS case. This suggests that adding disorder in this way might be sufficient to drive the system away from the Luttinger liquid, but the present data is clearly insufficient to draw any conclusion. To illustrate why it is risky to extrapolate from such small system sizes we consider a truncated version of (\ref{eq:hlongrange}) where all couplings for $|i-j|>3$ are removed. In that case adding disorder increases the fluctuations for small system sizes but not for large sizes, at this level of statistical resolution ($10^3$ realization of the disorder). Even though both scaling behaviors seem to follow a logarithmic scaling, it is not guaranteed that this remains correct when $L\to \infty$. For example a truncation to next nearest neighbors without disorder is known to be gapped with a very large but finite correlation length \cite{Okamoto}. 

 \section{Conclusion}
 \label{sec:conclusion}
 In this paper, we have studied the full counting statistics in ground states of spin chains of the Haldane-Shastry type. These wave functions can be seen, roughly speaking, as 1d discretizations of Laughlin states. They are also good variational Ans\"atze for ground states of spin chains in the LL universality class. 
 
 Our results may be summarized as follows. First, we provided a few finite-size formulae for correlations and the full (magnetization) counting statistics. These results were derived using random matrix theory techniques. Aside from free fermions, this provides another case were such exact formulae can be derived. We have also checked the scaling of fluctuations and the FCS as the system size is increased, both analytically and numerically. The results were in excellent agreement with the prediction of Luttinger liquid theory. In particular, the FCS diverges logarithmically with the subsystem size $\ell$, the prefactor being proportional to the Luttinger parameter, as well as the variable $\lambda^2$ where $\lambda$ is the counting parameter. This implies that only the second cumulant diverges, a clear signature of a gaussian effective theory in the continuum limit.
 
 Second, we showed that the aforementioned formulae can be generalized to inhomogeneous versions of our variational Jastrow-type state, where the particles need not be regularly spaced on the circle. The only additional requirement is the diagonalization/inversion of a $\sim L\times L$ matrix, which can be performed relatively fast for very large systems, using standard linear algebra routines. This result motivated us to revisit the problem of disorder in such states. We performed large scale simulations, averaged over many realizations of the disorder. Our results showed that disorder introduced in such a way is essentially irrelevant: the inhomogeneous versions of the Haldane-Shastry chain still lie in the Luttinger liquid universality class. The interpretation is that such disorder (preserving ``integrability'') is heavily correlated. This prevents any flow to the random singlet phase, notwistanding possible issues with the effect of long range interactions, which are currently not well understood. The situation is very different from the case of free fermions, which can easily be driven to the random singlet phase, without breaking integrability.
 
 A few issues are left as important open problems. The main one is that of disorder in systems with long range interactions. In our case the real space renormalization group treatment is not straightforward, and it is unclear whether the Haldane-Shastry chain can be truly disordered. We note also that the idea of moving the particles on the circle bears some similarity to certain Bethe Ansatz calculations, where inhomogeneities are introduced to make certain complicated expressions more tractable. Our results suggest that those inhomogeneities do not affect the long-distance properties of the system, for example the fact that the XXZ spin chain is in the Luttinger liquid universality class.
 
 Another intriguing question is that of the relation between fluctuations and entanglement. For free systems both can be computed in closed form, and there is a exact relation between the two \cite{KlichLevitov,Yalefluc1,Yalefluc2}. However the relation does not survive in interacting systems, as was confirmed numerically in fractional quantum Hall states \cite{Yalefluc3}. In the XXZ chain both fluctuations and entanglement are very difficult to compute using integrability techniques\cite{EE_XXZ}. 

 In the Haldane Shastry we have shown that the fluctuations can be computed exactly. However, an exact closed form expression for the entanglement is not known, even though we speculate that such a calculation might be possible.
 
 For clean system it would be interesting to determine whether our results for finite systems can be further generalized. For example, one could think of studying correlations of other observables such as $S^x$ or $S^y$. For two-point correlations these can easily be deduced using the $SU(2)$ symmetry; however higher order correlations should become more complicated. Another interesting direction would be to try and extend our results to variational states with higher symmetries, such as $SU(n)_1$ or $SU(2)_2$.
 
 \acknowledgments
 We thank Roberto Bondesan, Thomas Quella, Stephan Rachel, Nicolas Laflorencie, Gr\'egoire Misguich and Vincent Pasquier for stimulating discussions. Some of the DMRG calculations were performed using the ITensor C++ library.
 \bibliography{FCSHS}
\onecolumngrid
 \appendix
 \section{Correlations and fluctuations}
 \label{sec:app1}
 In this appendix we study the following variational state
\begin{equation}\label{eq:psil}
 \psi(\theta_1,\theta_2,\ldots,\theta_N)\propto \prod_{1\leq j<k\leq N} \left(e^{i \theta_j}-e^{i \theta_k}\right)^2 \propto  \prod_{1\leq j<k\leq N} \sin^2 \left(\frac{\theta_j-\theta_k}{2}\right)
\end{equation}
where $\theta_j=2\pi x_j/L$, and the $x_j$'s are a set of $L$ positions for the $N$ particles, in the range $[0,L]$.
 When the $x_j$ are integers and $N=L/2$, this is the ground state of the Haldane-Shastry chain, but we keep a general $N\leq L/2$ in the following (The case $N>L/2$ follows from the particle-hole symmetry). Since we are only interested in correlations which are diagonal in the particle basis, the extra phase factors which are present in (\ref{eq:psil}) can be safely discarded.  We will mostly use $\theta$ variables, so that e. g. the spin operator at site $x_j$ becomes $S_{\theta_j}^z$. Our main result will be a (square root) determinant formula for the diagonal correlations in the state (\ref{eq:psil}). Before going any further, let us note that the result is known in case $\theta$ is a continuous variable\cite{Mehta} -- this corresponds to the limit $L\to \infty$ with $N$ fixed-- or for discrete but regularly spaced ($x_j=j$) particles \cite{MehtaMehta74}. In the former case the correlations are essentially that of the circular ensemble for random matrices with a symplectic symmetry ($\beta=4$). We extend here the calculation to a case where the particles can only occupy $L$ different boxes with angles $\{\theta_j\;,\;j=1,\ldots,L\}$, not necessarily regularly spaced on the circle; it turns out this amounts only to minor complications. We also provide generalizations to open systems.

Our method will follow closely Refs.~\onlinecite{Gaudin,TracyWidom}, and uses only elementary techniques. The appendix is organized as follows. In appendix.~\ref{app:norm}, we explain how the norm of the state (\ref{eq:psil}) may be computed. In appendix.~\ref{app:correlations} we turn to the actual computation of the correlation functions. The results are then compared in appendix.~\ref{app:cft} to the predictions of conformal field theory. Finally, the case of open Haldane-Shastry chains is investigated in appendix.~\ref{sec:open}.

\subsection{Warm-up exercise: the normalization}
\label{app:norm}
Let us first consider the normalization $Z(L,N)$ of the state (\ref{eq:psil}). We have
\begin{equation}
 Z(L,N)=\frac{1}{N!}\sum_{\{\theta\}} \left|\psi(\theta_1,\ldots,\theta_N)\right|^2
\end{equation}
The $N!$ accounts for the fact that the positions of the particles/angles are now unordered. Now recall the following Vandermonde identity
\begin{equation}\label{eq:vand}
  V(z_1,z_2,\ldots,z_N)=\displaystyle{\det_{1\leq i,j\leq N}}\left(z_i^{j-1}\right)=
  \prod_{1\leq i<j\leq N} (z_j-z_i).
\end{equation}
What matters to compute the norm and the correlations is the square of $|\psi(\theta_1,\ldots,\theta_N)|$, which is the fourth power of the Vandermonde determinant with $z_j=e^{i\theta_j}$. Crucially, it may be obtained as the limit
\begin{equation}
 V(z_1,z_2,\ldots,z_N)^4=\lim_{\{w_i\}\to \{z_i\}} \frac{V(z_1,w_1,\ldots,z_N,w_N)}{(w_1-z_1)\ldots (w_N-z_N)}.
\end{equation}
By appropriate row-column manipulations on the determinant, one can then show
\begin{equation}\label{eq:confvand}
 V(z_1,z_2,\ldots,z_N)^4=\det_{\scriptsize{\begin{array}{c}1\leq j \leq N\\1\leq k \leq 2N\end{array}}} \left(\begin{array}{c}z_j^{k-1}\\(k-1)z_j^{k-2}\end{array}\right),
\end{equation}
an identity sometimes referred to as ``confluent Vandermonde'' \cite{Krattenthaler}. Using
\begin{equation}
 |e^{i\theta}-e^{i\phi}|=ie^{-i \frac{\theta+\phi}{2}}\left(e^{i\theta}-e^{i\phi}\right),
\end{equation}
we obtain
\begin{equation}
\left| \psi(\{\theta_j\})\right|^2=\frac{1}{2^N}\sum_{P\in S_{2N}} (-1)^P A_{p_1 p_2}^{\theta_1}A_{p_3 p_4}^{\theta_2}\ldots A_{p_{2N-1}p_{2N}}^{\theta_N},
\end{equation}
where the sum runs over all permutations of the half integers $(-N+1/2,-N+3/2,\ldots,N-1/2)$, $(-1)^P$ is the signature of the permutation, and
\begin{equation}
 A_{pq}^{\theta}=(q-p)e^{i\theta(p+q)}.
\end{equation}
Let us now perform the sum over the positions of the particles. We have
\begin{eqnarray}\nonumber
 Z&=&\frac{1}{N!}\sum_{\{\theta\}} \left|\psi(\theta_1,\ldots,\theta_N)\right|^2\\\nonumber
 &=&\frac{1}{2^N N!}\sum_{\{\theta\}}\sum_P (-1)^P A_{p_1 p_2}^{\theta_1}\ldots A_{p_{2N-1}p_{2N}}^{\theta_N}\\\label{eq:pf}
 &=&\frac{1}{2^N N!}\sum_P (-1)^P A_{p_1p_2}A_{p_3p_4}\ldots A_{p_{2N-1}p_{2N}}
\end{eqnarray}
with
\begin{equation}\label{eq:matelements}
 A_{pq}=\sum_\theta A_{pq}^{\theta}=(q-p)\sum_\theta e^{i\theta (p+q)}
\end{equation}
Eq.~(\ref{eq:pf}) is nothing but the Pfaffian of the antisymmetric matrix $A=(A_{pq})_{1\leq p,q\leq N}$,
\begin{equation}
Z(N,L)={\rm Pf\,} A. 
\end{equation}
This is quite useful, as the square of the Pfaffian is the determinant, and this makes numerical computations easy. For a clean system (corresponding to regularly spaced angles) the matrix elements (\ref{eq:matelements}) simplify greatly:
\begin{equation}\label{eq:apq}
 A_{pq}=(q-p)\sum_{\theta} e^{i\theta(p+q)}=(q-p)\sum_{x=1}^L e^{2i\pi (p+q)x/L}=L(q-p)\delta_{p,-q},
\end{equation}
 and $A$ is antidiagonal. The computation of the Pfaffian then trivializes, and we finally obtain
\begin{eqnarray}
 Z(L,N)
 &=&L^N(2N-1)!!,
\end{eqnarray}
where $!!$ denotes the double factorial, $(2N-1)!!=1\times 3\times 5\times \ldots \times (2N-1)$. This is an old result by Gaudin\cite{Gaudin}. 
\subsection{Correlations}
\label{app:correlations}
We have seen that the norm is given by
\begin{equation}
 Z(L,N)=\underset{p,q}{\rm Pf} \left(\sum_\theta A_{pq}^{\theta}\right).
\end{equation}
We wish to study $m$-point correlators, involving the sites $\theta_1,\ldots,\theta_m$.
To do that, let us introduce $\alpha(\theta)=\sum_{i=1}^m u_i \delta_{\theta,\theta_i}$ and consider the ratio
\begin{equation}\label{eq:cdef}
 C(u_1,\ldots,u_m)=\frac{ \underset{p,q}{\rm Pf} \left(\sum_\theta (1+\alpha(\theta))A_{pq}^{\theta}\right)}{ \underset{p,q}{\rm Pf} \left(\sum_\theta A_{pq}^{\theta}\right)}.
\end{equation}
It turns out $C(u_1,\ldots,u_m)$ generates all correlation functions relevant to our study. Say we are interested in the multipoint density correlations $\braket{\rho(\theta_1)\rho(\theta_2)\ldots \rho(\theta_m)}$, namely the joint probability that these sites be occupied by a particle (spin up here). 
One can check that the coefficient of $u_1 u_2\ldots u_m$ in $C(u_1,\ldots,u_m)$ is
\begin{equation}
\frac{\sum_{\theta_{m+1},\theta_{m+2},\ldots,\theta_N} \sum_P (-1)^P A_{p_1 p_2}^{\theta_1}\ldots A_{p_{2N-1}p_{2N}}^{\theta_N}
}{\sum_{\theta_{1},\theta_{2},\ldots,\theta_N} \sum_P (-1)^P A_{p_1 p_2}^{\theta_1}\ldots A_{p_{2N-1}p_{2N}}^{\theta_N}}
\;=\;\frac{\sum_{\theta_{m+1},\theta_{m+2},\ldots,\theta_N} |\psi(\theta_1,\ldots,\theta_N)|^2}{\sum_{\theta_{1},\theta_{2},\ldots,\theta_N} |\psi(\theta_1,\ldots,\theta_N)|^2}\,
,
\end{equation}
which is exactly $\braket{\rho(\theta_1)\ldots \rho(\theta_m)}$. Said differently,
\begin{equation}\label{eq:densitychoice}
 \braket{\rho(\theta_1)\ldots \rho(\theta_m)}=\frac{d}{du_1}\ldots \frac{d}{du_m}C(u_1,\ldots,u_m).
\end{equation}
Since $S^z_\theta=\rho(\theta)-1/2$,
 the spin correlations are given by
\begin{equation}\label{eq:spinchoice}
 \braket{S_{\theta_1}^z\ldots S_{\theta_m}^z}=(-2)^{-m}C(-2,\ldots,-2),
\end{equation}
and the FCS for the set of spins $\theta_1,\ldots,\theta_m$ is
\begin{equation}\label{eq:fcschoice}
 \Braket{e^{i\lambda \sum_{j=1}^m S_{\theta_j}^z}}=e^{-i\lambda m/2}C(e^{i\lambda}-1,\ldots,e^{i\lambda}-1).
\end{equation}
The expression (\ref{eq:cdef}) can be further simplified. Suppose we are able to invert the matrix $A=(A_{pq})$. In the clean case this is trivial, as $A$ is antidiagonal, but for the inhomogeneous case this can always be achieved numerically. With $\Gamma=A^{-1}$, and using ${\rm Pf}^2=\det$, we obtain
\begin{equation}\label{eq:density1}
 C(u_1,\ldots,u_m)^2=
\det_{p,q}\left(\delta_{pq}+\sum_{\theta}\sum_k \alpha(\theta) \Gamma_{pk}A_{kq}^{\theta}\right).
\end{equation}
The key observation is that the sum over $k$ in the determinant can be rewritten as a scalar product
\begin{equation}\label{eq:sometrick}
 L^2\sum_k \Gamma_{pk}A_{pq}^{\theta}=\left(\begin{array}{cc}i\sum_k \Gamma_{pk} ke^{i k\theta}&\sum_k \Gamma_{pk}e^{i k\theta}\end{array}\right)
 \left(\begin{array}{c}ie^{iq\theta}\\qe^{iq\theta}\end{array}\right)
\end{equation}
Using (\ref{eq:sometrick}), (\ref{eq:density1}) becomes
\begin{equation}\label{eq:density2}
C(u_1,\ldots,u_m)^2=\det\left(1+PQ\right)
\end{equation}
where $P$ is a $2N\times 2L$ matrix, and $Q$ is a $2L\times 2N$ matrix. Even though these are rectangular, the product is square and the identity $\det(1+PQ)=\det(1+QP)$ holds. 
$QP$ is a $L\times L$ block matrix, given by
\begin{equation}
 QP=\left(
 \begin{array}{cccc}
  \alpha(\phi)\frac{\partial f(\theta,\phi)}{\partial \phi}
  &&\alpha(\phi)f(\theta,\phi)\\\\
  -\alpha(\phi)\frac{\partial^2 f(\theta,\phi)}{\partial \theta\partial \phi}&&-\alpha(\phi)\frac{\partial f(\theta,\phi)}{\partial \theta}
 \end{array}
 \right)_{\footnotesize{\textrm{$\begin{array}{l}\theta=\theta_1,\ldots,\theta_L\\ \phi=\phi_1,\ldots,\phi_L\end{array}$}}},
\end{equation}
with
\begin{equation}\label{eq:f}
 f(\theta,\phi)=\frac{i}{L^2}\sum_{p,k}\Gamma_{pk}e^{i(p\theta+k\phi)} \qquad,\qquad \Gamma=A^{-1}.
\end{equation}
Notice that since $A$ is antisymmetric, so is $\Gamma$, and $f(\theta,\phi)=-f(\phi,\theta)$. 
The point of this manipulation is that now all columns in $QP$ are proportional to $\alpha(\phi)=\sum_{i=1}^m u_i \delta_{\phi,\theta_i}$, and so each column corresponding to a site not in $\{\theta_1,\ldots,\theta_m\}$ is identically zero. 
Hence, $C(u_1,\ldots,u_m)$ reduces to the square-root of a smaller $2m\times 2m$ determinant:
\begin{equation}
 C(u_1,\ldots,u_m)^2=\det_{1\leq i,j\leq m}\left[\left(\begin{array}{cc}\delta_{ij}&0\\0&\delta_{ij}\end{array}\right)
 +u_j K(\theta_i,\theta_j)
 \right],
\end{equation}
with a kernel
\begin{equation}\label{eq:R}
 K(\theta,\phi)=\left(
 \begin{array}{ccc}
  \frac{\partial f(\theta,\phi)}{\partial \phi}&&f(\theta,\phi)\\\\
 -\frac{\partial^2 f(\theta,\phi)}{\partial \theta\partial \phi} && -\frac{\partial f(\theta,\phi)}{\partial \theta}
 \end{array}
 \right).
\end{equation}
The spin correlations and FCS are then recovered using Eqs.~(\ref{eq:spinchoice},\ref{eq:fcschoice}).
All these results can alternatively be written as Pfaffians. For example one gets
\begin{equation}
  \braket{S_{\theta_1}^z\ldots S_{\theta_m}^z}=\underset{1\leq i,j\leq m}{\rm Pf}
  \left(\begin{array}{cc}
         f(\theta_i,\theta_j)&\frac{\partial f(\theta_i,\theta_j)}{\partial \theta_i}+\frac{\delta_{ij}}{2}\\\\
         \frac{\partial f(\theta_i,\theta_j)}{\partial \theta_j}-\frac{\delta_{ij}}{2}&\frac{\partial^2 f(\theta_i,\theta_j)}{\partial \theta_i \partial \theta_j}
        \end{array}
  \right)
\end{equation}
for the $m$ point function.
\subsection{Clean case, and connection with conformal field theory}
\label{app:cft}
As was already mentionned, for regularly spaced particles on the circle $A$ is antidiagonal, $A_{pq}=\frac{q-p}{L}\delta_{p,-q}$, so $\Gamma_{pq}=(A^{-1})_{pq}=-\frac{L}{q-p}\delta_{p,-q}$ and the ancillary function (\ref{eq:f}) can be computed explicitely. We obtain
\begin{equation}
 f(\theta,\phi)=\frac{1}{L}\sum_{p>0} \frac{\sin p(\theta-\phi)}{p},
\end{equation}
where the sum runs over the half integers $p=1/2,3/2,\ldots,N-1/2$. 
From this all derivatives can easily be obtained.
The result can be recast as a Pfaffian:
\begin{equation}\label{eq:pfaffianresult}
 \Braket{S_{\theta_1}^z\ldots S_{\theta_m}^z}=\underset{1\leq i,j\leq m}{\rm Pf}\left(G_N(\theta_i,\theta_j)\right)
\end{equation}
where the $2\times 2$ kernel is given by
\begin{equation}\label{eq:pfaffiankernel}
 G_N(\theta,\phi)=\left(\begin{array}{ccc}
               a_N(\theta-\phi)&&f_N(\theta-\phi)-\delta_{\theta,\phi}/2\\
               \delta_{\theta,\phi}/2-f_N(\theta-\phi)&&d_N(\theta-\phi)
              \end{array}
 \right)
\end{equation}
$f_N(\theta)$ is given by
\begin{equation}
 f_N(\theta)=\frac{\sin N\theta}{2L\sin \frac{\theta}{2}}
\end{equation}
Note $f_N(0)=N/L$. The two others are
\begin{eqnarray}\label{eq:abis}
 a_N(\theta)&=&\int_0^\theta f_N(\phi)d\phi=\frac{1}{L}\sum_{k=1}^N \frac{\sin \left[(k-1/2)\theta\right]}{k-1/2}
 \\\label{eq:d}
 d_N(\theta)&=&\frac{df_N(\theta)}{d\theta}=\frac{2N \cos N\theta-\cot\frac{\theta}{2} \sin N\theta}{4L \sin \frac{\theta}{2}}
\end{eqnarray}
and are antisymmetric, so that the Pfaffian makes sense.
Note that from our analysis
\begin{equation}
 \braket{S_{\theta_1}^z}=N/L-1/2
\end{equation}
which is obvious due to translational invariance. Nice simplifications occur at half-filling $N=L/2$ because half the matrix elements in (\ref{eq:pfaffianresult}) are now zero. It is then possible to reorganize the $a_N$ and $d_N$ separately, and get a product of two smaller Pfaffians
\begin{equation}\label{eq:hscorr}
\Braket{S_{x_1}^z S_{x_2}^z\ldots S_{x_m}^z}\;=\;\left[\underset{1\leq i,j\leq m}{\rm Pf}\left(\frac{(-1)^{x_i-x_j}}{\frac{L}{\pi}\sin \frac{\pi}{L}(x_i-x_j)}\right)\right]\;\times\;
\left[\underset{1\leq i,j\leq m}{\rm Pf}\left(\sum_{k=1}^{L/2}\frac{\sin\left[ \frac{(2k-1)\pi (x_i-x_j)}{L}\right]}{2\pi(2k-1)}\right)\right]
\end{equation}
where we have put back the positions of the spins, $\theta=2x\pi/L$. Specifying $m=2$ in (\ref{eq:hscorr}) reproduces the result of Ref.~\onlinecite{BBG} (see also Refs.~\onlinecite{NielsenCiracSierra,StephanSMI}),
\begin{equation}
 \Braket{S_{x_1}^zS_{x_2}^z}=\frac{(-1)^{x_1-x_2}}{2L\sin \frac{\pi}{L}(x_1-x_2)}\sum_{k=1}^{L/2}\frac{\sin\left[ \frac{(2k-1)\pi (x_1-x_2)}{L}\right]}{2k-1}.
\end{equation}
For odd $m$ the Pfaffian always gives zero, consistent with the fact that the ground-state is in the sector with zero magnetization.
Now let us check that the long distance limit $|x_i-x_j|\gg 1$, $L\gg 1$ is consistent with known conformal field theory (CFT) results. It turns out the first Pfaffian on the left in (\ref{eq:hscorr}) is \emph{exactly} the CFT correlator of vertex operators for $SU(2)_1$. All finite-size effect are therefore encoded in the second Pfaffian. Indeed, one can check that
\begin{equation}
 \sum_{k=1}^{L/2}\frac{\sin\left[ \frac{(2k-1)\pi x}{L}\right]}{\pi(k-1/2)}=\frac{{\rm sgn}\; x}{2}-\frac{(-1)^x}{\frac{L}{\pi}\sin \frac{\pi x}{L}}+O(1/L^2)
\end{equation}
so the second Pfaffian only contributes a constant term to the leading order in the correlator. (We have used ${\rm Pf}\,[ {\rm sgn}(j-i)]=1$). Another interesting limit is that of an infinite system, in which case we obtain
\begin{equation}
 \Braket{S_{x_1}^z S_{x_2}^z\ldots S_{x_m}^z}\;=\;
 \left[\underset{1\leq i,j\leq m}{\rm Pf}\left(\frac{(-1)^{x_i-x_j}}{x_i-x_j}\right)\right]
 \times \left[\underset{1\leq i,j\leq m}{\rm Pf}\left(\frac{{\rm Si}\,(\pi[x_i-x_j])}{4\pi}\right)\right].
\end{equation}
${\rm Si}$ denotes the integral sine, ${\rm Si}\;u=\int_0^u \frac{\sin t}{t}$.

Note that such a decoupling as a product of two $m\times m$ Pfaffians does not seem to occur away from half-filling. Instead we keep a bigger $2m\times 2m$ Pfaffian similar to (\ref{eq:pfaffianresult}).
\subsection{Open systems}\label{sec:open}
Similar results can also be established for the ground-state of the HS chain with open boundary conditions\cite{SimonsAltshuler,BernardPasquierSerban,TuSierra}. We study the (unnormalized) state
\begin{equation}\label{eq:openstate}
 \psi(\theta_1,\ldots,\theta_N)=\prod_{j=1}^N \sin^{\alpha+1} \theta_j \prod_{k=j+1}^N \left(\cos \theta_j-\cos \theta_k\right)^2,
\end{equation}
where the $\theta_j$ are now a set of $N$ angles in $(0;\pi)$, and $\alpha$ is left as a generic parameter for now. This amplitude may be rewritten as
\begin{equation}
 \psi(\theta_1,\ldots,\theta_N)\propto \prod_{j=1}^N \left(1-e^{2i\theta_j}\right)^{\alpha+1}\prod_{k=j+1}^{N} \left(e^{i\theta_j}-e^{i \theta_k}\right)^2 \left(1-e^{i(\theta_j+\theta_k)}\right)^2
\end{equation}
For correlations in the particle basis what matters is once again the square of this amplitude. To treat it we make use of the identities
\begin{equation}
 \det_{1\leq j,k\leq N} \left(z_j^k-z_j^{-k}\right)=\prod_{j=1}^N z_j^{-N}\left(1-z_j^2\right)\prod_{k=j+1}^N
 \left(z_j-z_k\right)
 \left(1-z_j z_k\right),
\end{equation}
and
\begin{equation}
\det_{\scriptsize{\begin{array}{c}1\leq j \leq N\\1\leq k \leq 2N\end{array}}}
 \left(\begin{array}{c} z_j^k-z_j^{-k}\\k z_j^{k-1}+k z_j^{-k-1}\end{array}\right)=\left[\det_{1\leq j,k\leq N}\left(z_j^k-z_j^{-k}\right)\right]^4 \prod_{j=1}^N \left(z_j^2-1\right)^{-1}
 ,
\end{equation}
which may be proven in similar fashion as their periodic counterparts (\ref{eq:vand}) and (\ref{eq:confvand}). We find that the norm of the state (\ref{eq:openstate}) is
\begin{equation}
 \frac{1}{N!}\sum_{\{\theta\}}\psi(\theta_1,\ldots,\theta_N)^2\;=\;\underset{1\leq p,q\leq 2N}{\rm Pf}\left(A_{pq}\right)
\end{equation}
where
\begin{equation}
 A_{pq}=\sum_\theta A_{pq}^\theta,
\end{equation}
and 
\begin{align}\nonumber
 A_{pq}^\theta&=\frac{(\sin \theta)^{2\alpha-1}}{2^{2n-1}}
 \left[p \cos p\theta \sin q \theta-q \sin p \theta \cos q\theta\right]
 \\
 &=\frac{(\sin \theta)^{2\alpha-1}}{2^{2n}}
 \left[
 (p-q)\sin (p+q)\theta-(p+q)\sin (p-q)\theta
 \right].
\end{align}
The generating function $C(u_1,\ldots,u_m)$ is now given, in block form, by
\begin{equation}
 C(u_1,\ldots,u_m)^2=\det_{1\leq i,j\leq m}\left[\left(\begin{array}{cc}\delta_{ij}&0\\0&\delta_{ij}\end{array}\right)
 +u_j K(\theta_i,\theta_j)
 \right],
\end{equation}
with a kernel
\begin{equation}
 K(\theta,\phi)=(\sin \phi)^{2\alpha-1}\left(
 \begin{array}{cc}-\frac{\partial g}{\partial \phi}&-g\\\frac{\partial^2 g}{\partial \theta \partial \phi}& \frac{\partial g}{\partial \theta}\end{array}
 \right)
 \qquad,\qquad g(\theta,\phi)=\sum_{p=1}^{2N}\sum_{k=1}^{2N}\Gamma_{kp}\sin p\theta \sin k \phi\qquad,\qquad \Gamma=A^{-1}.
\end{equation}
The correlations of interest are then reconstructed as
\begin{eqnarray}
 \Braket{\rho(\theta_1\ldots\theta_m)}&=&\frac{d}{du_1}\ldots \frac{d}{du_m}C(u_1,\ldots,u_m)=\left[\det_{1\leq i,j\leq m}\left(K(\theta_i,\theta_j)\right)\right]^{1/2},\\
 \Braket{S_{\theta_1}^z\ldots S_{\theta_m}^z}&=&(-2)^{-m}C(-2,\ldots,-2),\\
 \Braket{e^{i\lambda \sum_{j=1}^m S_{\theta_j}^z}}&=&e^{-im\lambda/2}C(e^{i\lambda}-1,\ldots,e^{i\lambda}-1).
\end{eqnarray}
All these can also be rewritten as Pfaffians. 
Of course, such expressions are only fully explicit if the matrix $A$ can be inverted exactly. This is the case for 
regularly spaced angles $\theta_j$ and some simple integer values of $\alpha$. For example when $\alpha=0$, all  three cases $\theta_j=\frac{(j-1/2)\pi}{L}$, $\theta_j=\frac{j\pi}{L+1}$, $\theta_j=\frac{2j\pi}{2L+1}$ for $j\in \{1,\ldots,L\}$, lead to a matrix $A_{pq}\propto \sin^2 \frac{\pi(p-q)}{2}\left[p-q-(p+q)\,{\rm sgn}(p-q)\right]$ which can be inverted. These are respectively dubbed open Haldane-Shastry chains of type (I), (II) and (III) in Ref.~\onlinecite{BernardPasquierSerban}. Our formula also reproduces the result of Ref.~\onlinecite{TuSierra} when specified to the type (I) two-point function. The value $\alpha=1$ is also interesting, as the wave function amplitude turns out (in case II) to be exactly the square of the ground state amplitude of the XX chain with open boundary conditions, which lies in the same universality class. For all three types we get a matrix $A_{pq}\propto (p+q)\left[\delta_{p,q+1}-\delta_{p+1,q}\right]$ (up to an extra boundary term for type (I) and $N=L/2$) which can also be inverted, leading to explicit expressions for all correlations.
\end{document}